\documentclass[prl, showpacs,twocolumn,superscriptaddress,floatfix,amsmath]{revtex4}
\usepackage[english]{babel}
\usepackage{graphicx}
\usepackage{amsmath}
\usepackage{amssymb}

\begin{document}
\title{Aharonov-Bohm oscillations in disordered topological insulator nanowires}


\author{J.~H.~Bardarson}
\affiliation{Department of Physics, University of California, Berkeley, Berkeley, California 94720, USA}
\affiliation{Materials Sciences Division, Lawrence Berkeley National Laboratory, Berkeley, California 94720, USA}
\author{P.~W.~Brouwer}
\affiliation{Dahlem Center for Complex Quantum Systems and Institut f\"ur
Theoretische Physik, Freie Universit\"at Berlin, Arnimallee 14, 14195 Berlin,
Germany}
\author{J.~E.~Moore}
\affiliation{Department of Physics, University of California, Berkeley, Berkeley, CA 94720}
\affiliation{Materials Sciences Division, Lawrence Berkeley National Laboratory, Berkeley, CA 94720}

\date{May 2010}
\begin{abstract}
A direct signature of electron transport at the metallic surface of a topological insulator is the Aharonov-Bohm oscillation
observed in a recent study of Bi$_2$Se$_3$ nanowires [Peng {\it et~al.}, Nature Mater.\ {\bf 9}, 225 (2010)] where conductance was found to oscillate as a
function of magnetic flux $\phi$ through the wire, with a period of one flux quantum $\phi_0 = h / e$ and {\it maximum} conductance
at zero flux.  This seemingly agrees neither with diffusive theory, which would predict a period of half a flux quantum, nor with ballistic theory, which in the simplest form predicts a period of $\phi_0$ but a {\it minimum} at zero flux due to a nontrivial Berry phase in topological insulators.
We show how $h/e$ and $h/2e$ flux oscillations of the conductance depend on doping and disorder strength, provide a possible explanation for the experiments, and discuss further experiments that could verify the theory.
\end{abstract}

\maketitle{}

The characteristic feature of a strong 3D topological insulator (TI) is the presence of a conducting surface that is topologically protected
from Anderson localization by time-reversal-invariant disorder~\cite{TIreview}. In general the surface state has an odd number of Dirac points in
the energy spectrum, with the
simplest case of a single Dirac point being realized at the 
(111) surface of Bi$_2$Se$_3$~\cite{Xia09,HZha09}. While the presence of this surface metallic state has been
demonstrated convincingly using surface probes, notably angle-resolved photoemission spectroscopy~\cite{Xia09}, studies of the transport properties of these
surfaces are rare~\cite{Tas09, Che09, Pen10, Che10,Ana10}. Topological insulator nanowire with perfectly insulating bulk realizes an
ideal hollow metallic cylinder with a diameter large enough that it is easy to thread a large magnetic flux through its core~\cite{Pen10}. The
magnetoconductance of such wires not only reflects the fundamental effects of normal metal physics such as the Aharonov-Bohm effect and weak
localization, but can also indirectly probe the existence of a nontrivial Berry phase.

\begin{figure}[tb!]
  \begin{center}
    \includegraphics[width=0.95\columnwidth]{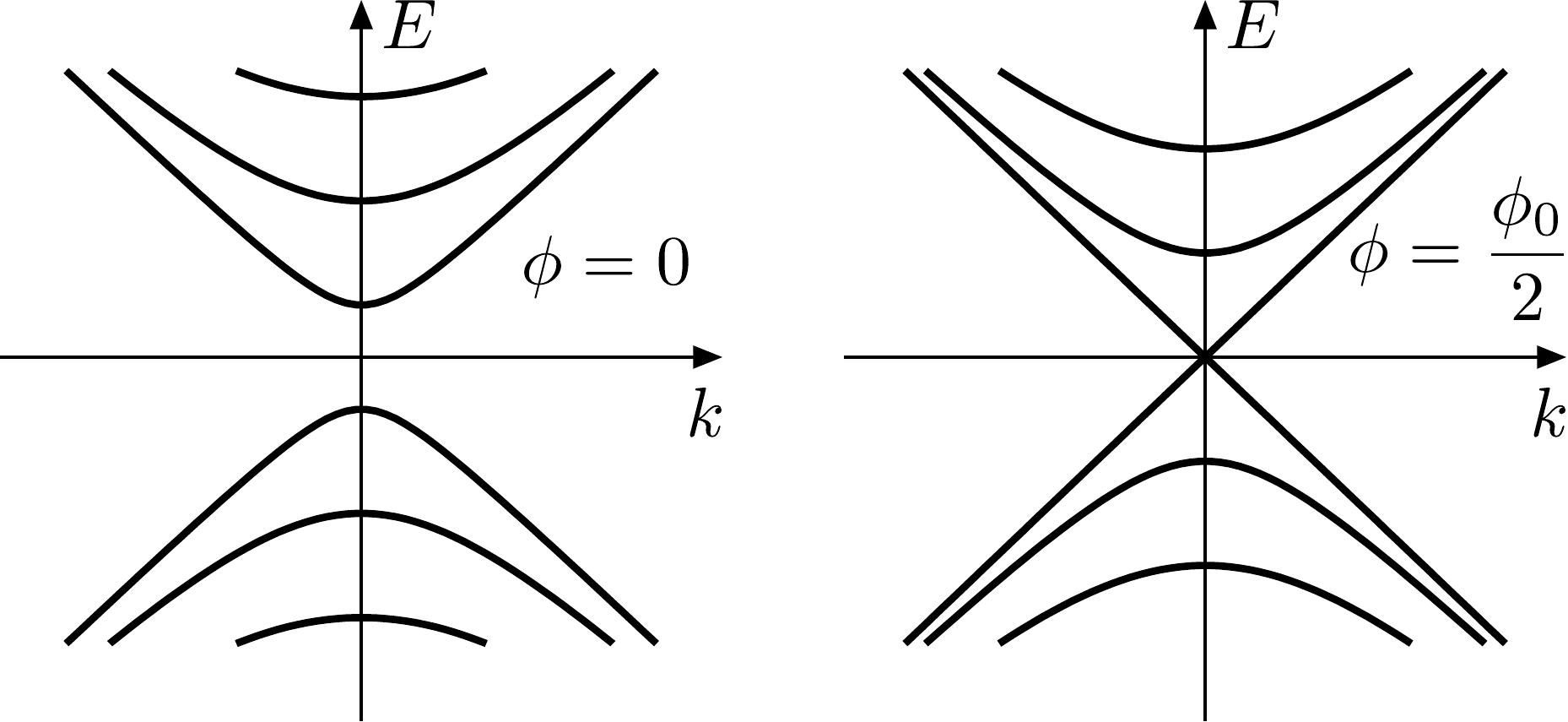}
  \end{center}
  \vspace{-0.5cm}
  \caption{A schematic of the band structure of a topological insulator nanowire in the presence of a magnetic flux $\phi$. Because of a nontrivial Berry
  phase of the particle a gapless mode is only realized at a flux with half integral flux quanta.}
  \label{fig:Bands}
\vspace{-0.5cm}
\end{figure}

The magnetic flux affects the transport properties of the metal surface through the Aharonov-Bohm effect: the wave function of the particles picks up a
phase of $2\pi \phi/\phi_0$ going around the circumference, with $\phi$ the total flux through the cylinder and $\phi_0 = h/e$ the flux
quantum~\cite{Aro87}. There are two
inequivalent values of flux that do not break time-reversal symmetry in the surface: 0 and $\phi_0/2$ (up to integer multiples of $\phi_0$). In normal metals there is no fundamental difference
between the two values, but in TI nanowires there is: only one of these values allows for a state at the Dirac point (cf.\
Fig.\ref{fig:Bands}). When there is a state at the Dirac
point, the total number of modes is odd and time-reversal symmetry requires the presence of a single perfectly transmitted
mode~\cite{And02} contributing conductance $e^2/h$. If the
contribution of all other modes to the conductance is exponentially suppressed, which happens for example at the Dirac point for ballistic
wires with a small aspect ratio (length $\gg$ circumference), the conductance is dominated by the presence or absence of the perfectly transmitted mode. In a flat surface with a
Dirac point, such as graphene, the zero mode is realized at zero flux, while in a curved surface as in TI nanowires, it is realized at
$\phi_0/2$~\cite{Ran08,Ost10,Rose10}. This phase shift occurs because the particle spin is constrained to lie in the tangent plane to the
surface and thus a particle
picks up a Berry phase of $\pi$ due to the $2\pi$ rotation of the spin as it goes around the surface~\cite{YZha09}. The magnetoconductance of an undoped ballistic TI
nanowire is thus expected to oscillate with a period of $\phi_0$ and a {\it maximum} at $\phi = \phi_0/2$.

The situation is different in the presence of disorder strong enough that transport is diffusive. The classical conductance acquires a quantum correction due to interference between time reversed paths. The phase difference between a path going once clockwise around the
cylinder and one going anticlockwise is $2\times 2\pi \phi/\phi_0$ and the conductance thus oscillates with a period of $\phi_0/2$~\cite{Aro87}. Whether
the conductance has a maximum or a minimum at zero flux depends on the presence (weak antilocalization) or absence (weak localization) of
spin-orbit coupling.  By their very nature, topological insulators have strong spin-orbit coupling and thus show weak antilocalization with a
maximum conductance at zero flux. 

Surprisingly, however, in the recent transport experiments on Bi$_2$Se$_3$ nanowires neither of the above scenarios seems to be
realized~\cite{Pen10}.
In the experiment the weak antilocalization induced $\phi_0/2$ period is essentially absent, while the $\phi_0$ periodicity is clearly seen. However,
the conductance has a maximum at $\phi = 0$ rather than at the expected $\phi = \phi_0/2$.
Although none of the available TI are particularly good bulk
insulators, an explanation of this discrepancy from bulk properties is unlikely, since there is no compelling reason
for the bulk conductance to show sharp flux periodicity determined by the cross-sectional area.

In this paper we provide a theoretical study of the transport properties of the surface in the presence of parallel flux and time-reversal preserving disorder, combining analytical estimates with numerical simulation. By studying a pure surface theory our results are not
complicated by bulk contributions, making it easier to disentangle the surface and bulk properties in the experiments. We show that in the
regime of weak disorder and nonzero doping our theory is consistent with the experimental result. We begin by briefly explaining our theoretical model and calculations before presenting our results and discussing them in relation to experiments.

The transport properties of the surface are determined by the Dirac equation
\begin{equation}
  \left[ v\mathbf{p}\cdot\mathbf{\sigma} + V(\mathbf{r}) \right]\psi = \varepsilon \psi.
  \label{eq:Dirac}
\end{equation}
$\mathbf{\sigma} = (\sigma_x, \sigma_y)$ are the Pauli sigma matrices and $v$ is the Fermi velocity. The Fermi energy $\varepsilon$ is determined
by the density of surface charge carriers and can in principle be tuned by doping. 
$V(\mathbf{r})$ is the
disorder potential which we take to be Gaussian correlated
\begin{equation}
  \langle V(\mathbf{r}) V(\mathbf{r}') \rangle =
   K_0\frac{(\hbar v)^2}{2\pi \xi^2} e^{-|\mathbf{r} - \mathbf{r}'|^2/2\xi^2}
  \label{eq:GaussCorr}
\end{equation}
with correlation length $\xi$. The exact form of the correlator is not important in obtaining our results. $K_0$ is the dimensionless measure of the disorder strength. We take $0<x<L$ to be the coordinate along the wire and
$0<y<W$ as the circumferential coordinate. The magnetic flux $\phi$ is absorbed into the
boundary condition
\begin{equation}
  \psi(x,y+W) = \psi(x,y)e^{i(2\pi\phi/\phi_0 + \pi)}.
  \label{eq:BC}
\end{equation}
The extra factor of $\pi$ is the curvature-induced Berry phase. 
Despite the large $g$ factor the Zeeman coupling of the magnetic field and the spin is not expected to be of relevance. Since the field
is parallel to the surface this simply shifts the band structure in Fig.~\ref{fig:Bands}, unlike a normal Zeeman field, which opens up a gap.
We therefore ignore Zeeman coupling in this work

\begin{figure}[t]
  \begin{center}
    \includegraphics[width=0.9\columnwidth]{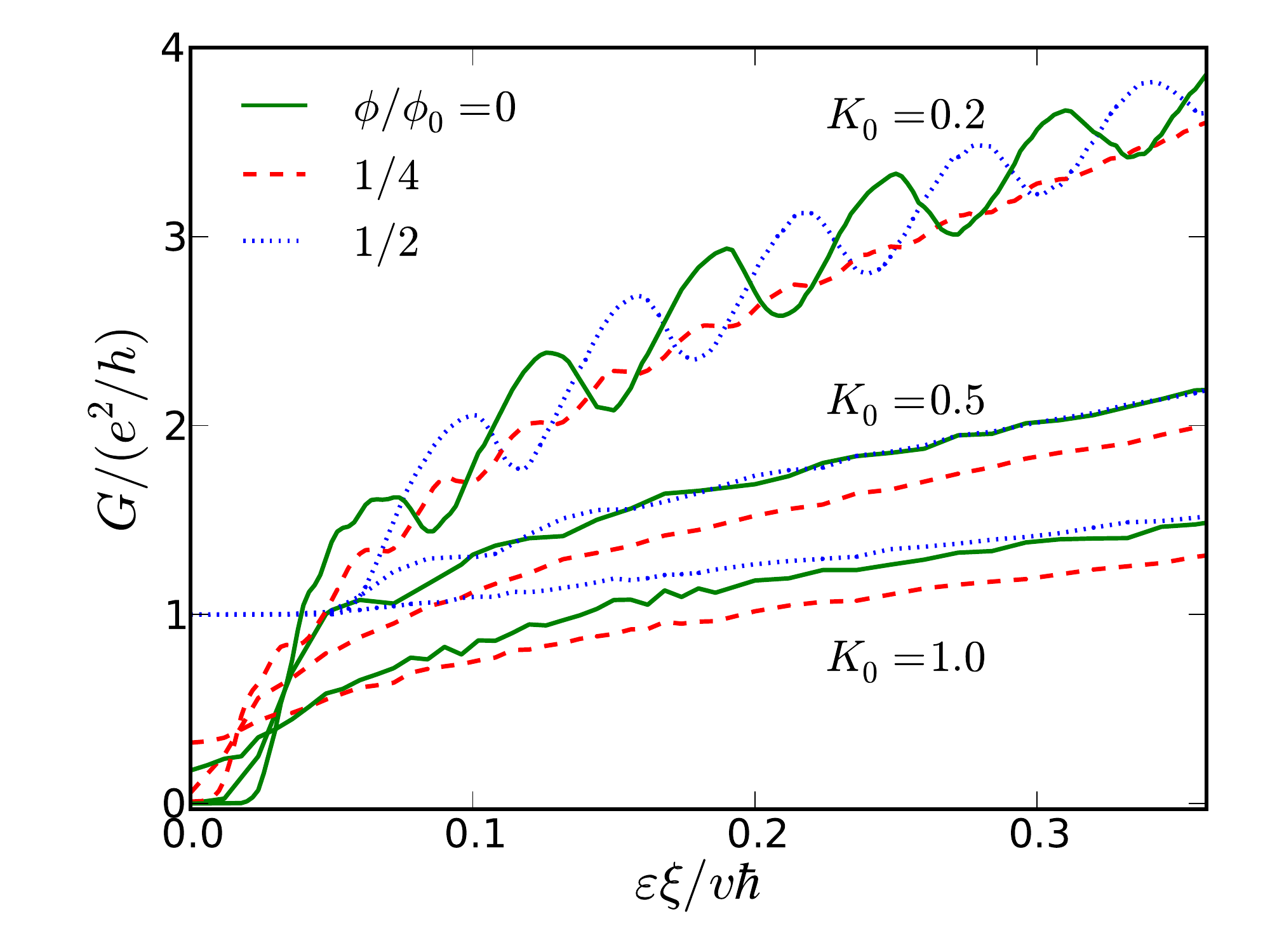}
  \end{center}
  \vspace{-0.5cm}
  \caption{Conductance versus Fermi energy for three different disorder strengths $K_0$ and for three representative values of magnetic flux.
  Here $W = 100\xi$ and $L=200\xi$.
  At low disorder (K=0.2), whether the conductance at $\phi=0$ or $\phi = \phi_0/2$ is larger is highly sensitive to the location of the Fermi level.}
  \label{fig:EnergyDependence}
  \vspace{-0.3cm}
\end{figure}

\begin{figure}[bt]
  \begin{center}
    \includegraphics[width=0.9\columnwidth]{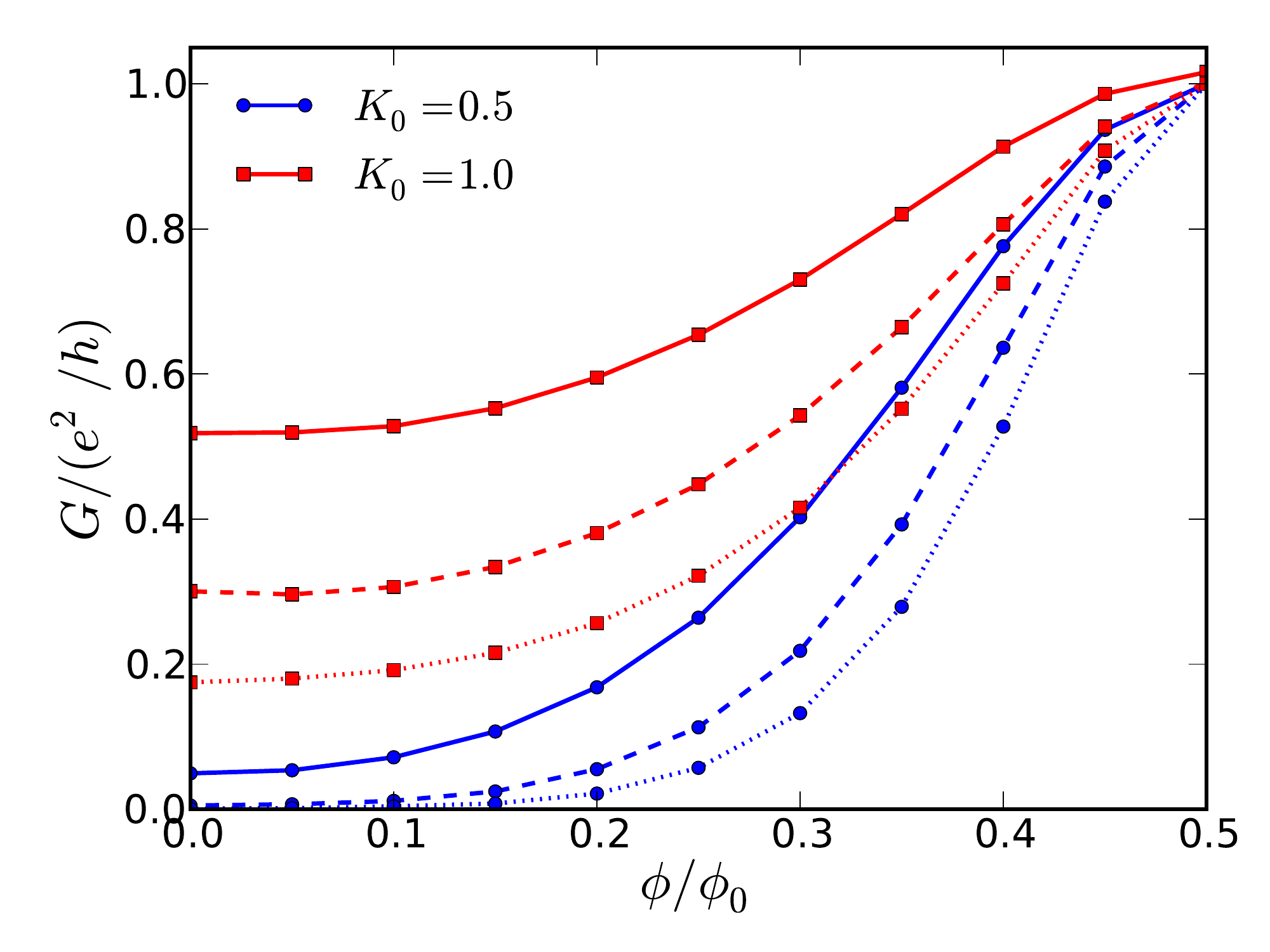}
  \end{center}
  \vspace{-0.5cm}
  \caption{Conductance at the Dirac point $\varepsilon = 0$ as a function of magnetic flux with $W=100\xi$, $L/\xi = 100$ (solid lines), $150$ (dashed lines) and $200$ (dotted
  lines), and two values for the disorder strength.}
  \label{fig:phiDep}
  \vspace{-0.4cm}
\end{figure}

The scattering matrix is obtained using the transfer matrix
method of Ref.~\onlinecite{Bar07}, which in turn gives the conductance $G$ through the Landauer formula. The total transfer matrix
$\mathcal{T}$ relates the wave function on the left ($x = 0$) to the wave function on the right ($x=L$) and is obtained as a product of $N$
transfer matrices $\delta \mathcal{T}_j$ that propagate the wave function from $x=(j-1)L/N$ to $x = jL/N$ with $j=1,\dots,N$. The $\delta
\mathcal{T}_j$ are obtained by solving the Dirac equation~\eqref{eq:Dirac}. The dimension of $\delta \mathcal{T}_j$ is determined by the
number $2M+1$ of transverse modes $q_n = 2\pi(n+\phi/\phi_0+1/2)/W,
n = -M, M+1,\dots,M$ included in the calculation. We take $N$ and $M$
large enough that the conductance no longer depends on them. For explicit expressions for the matrices $\delta \mathcal{T}_j$ and further
details we refer to Ref.~\onlinecite{Bar07}. We average over $\sim 10^3$ disorder
realizations.

The model~\eqref{eq:Dirac} has been studied extensively in the limit of large aspect ratio $W/L \gg 1$, in the context of graphene in the
presence
of valley-preserving scalar disorder (for a recent review see Ref.~\onlinecite{Das10}). In this limit the conductance is independent of boundary condition, and therefore of flux, and is
topologically protected from Anderson localization, growing logarithmically with system size indicative of weak
antilocalization~\cite{Bar07,Nom07}. The
opposite limit of very small aspect ratio $ W/L \ll 1$ describes carbon nanotubes if in addition $W$ is very small (smaller
than the mean free path). Carbon nanotubes are fundamentally different than TI wires in that the Berry phase is absent. Also, due to their small
diameter it is challenging to thread through them a flux of the order of or larger than a flux quantum. For a review of related theoretical
and experimental work see
Ref.~\onlinecite{Cha07}. Here we study the intermediate regime of aspect ratio $W/L \sim 1$, which is the regime relevant to the experiment of Ref.~\onlinecite{Pen10}. In this regime
the conductance does depend on the flux and this dependence in the presence of disorder has to our knowledge not been studied systematically before.

We now present and interpret the results of our numerical simulation.  In Fig.~\ref{fig:EnergyDependence} we plot the density dependence of the conductance for a few different disorder strengths and a few values for the magnetic flux. Three qualitatively different regimes are observed: (i) The small doping regime close to the Dirac point, and large doping regime with (ii) weak disorder (small $K_0$) and (iii) large disorder (large $K_0$).

In the small doping regime, the conductance is generally small compared to the conductance quantum $e^2/h$, and the physics in this regime is
thus dominated by the possibility of a perfectly transmitted mode. In Fig.~\ref{fig:phiDep} we show the flux dependence of the conductance
for a fixed diameter and a few wire lengths. Because of the Berry phase the conductance peaks at $\phi = \phi_0/2$ with
conductance of about $e^2/h$. For other values of flux the conductance is exponentially suppressed (with aspect ratio) and goes to zero
in the limit of $W/L \rightarrow 0$.  Interestingly, the conductance increases with increasing disorder strength for a fixed system size, at
the same time going to zero with increasing system size.  

\begin{figure}[t]
  \begin{center}
    \includegraphics[width=0.9\columnwidth]{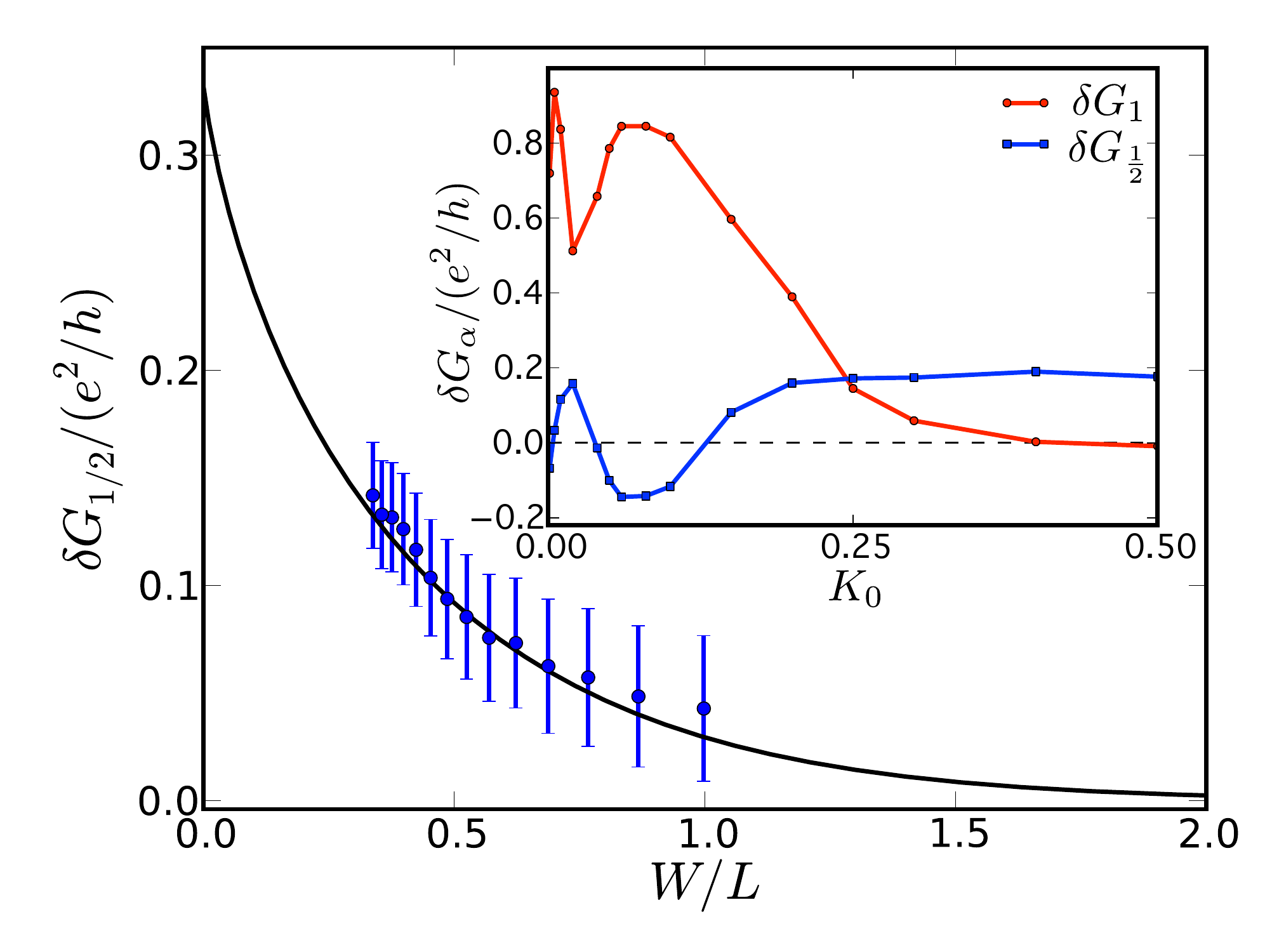}
  \end{center}
  \vspace{-0.6cm}
  \caption{Main panel: Comparison of the analytical expression~\eqref{eq:walG} for the weak antilocalization correction as a function of
  aspect ratio, with numerical results. The
  numerical data are obtained for a fixed $W = 40\xi$, $K_0 = 1$ and $\varepsilon\xi/v\hbar = 3$. Inset: The conductance amplitudes $\delta
  G_{1/2}$ and $\delta G_1$ of Eq.~\eqref{eq:DeltaG} as a function of disorder strength $K_0$ at $\varepsilon\xi/v\hbar =
  0.308$ and for $L=2W=200\xi$.  The structures at small $K_0$ can be understood as Fabry-Perot resonances.  As disorder increases, the half-flux oscillations dominate the one-flux oscillations. }
  \vspace{-0.4cm}
  \label{fig:K0dep}
\end{figure}

For weak disorder, away from the Dirac point, the magnetoconductance generically oscillates with a period of $\phi_0$ with an amplitude
that depends on $\varepsilon$. The maximum conductance is either at $\phi = 0$ or at $\phi = \phi_0/2$ depending on the amount of doping.
(Without doping, the maximum is at $\phi = \phi_0/2$, as discussed in the introduction.) 
In the inset to Fig.~\ref{fig:K0dep} we plot the amplitudes $\delta G_{1/2}$
and $\delta G_{1}$ of the $\phi_0/2$ and $\phi_0$-periodic oscillations of
the magnetoconductance,
\begin{subequations}
\begin{align}
  \delta G_{\frac{1}{2}} &= \frac{G(\phi=\phi_0/2) + G(\phi = 0) - 2G(\phi = \phi_0/4)}{2},\\
  \delta G_1 &= G(\phi = 0) - G(\phi = \phi/2),
\end{align}
\label{eq:DeltaG}
\end{subequations}
as a function of disorder strength for $\epsilon = 0.308 \hbar v/\xi$ as an example of a doping level where the conductance has a maximum at
$\phi = 0$ (as in the experiment~\cite{Pen10}). $\delta G_{\alpha}$ is a measure of the strength of the $\alpha\phi_0$ period in the
magnetoconductance. 

For very small values of $K_0$ the conductance can have sharp Fabry-Perot resonances that leads to the complicated
nonmonotonic behavior seen in Fig.~\ref{fig:K0dep}. The resonances are very sensitive to disorder and disappear already at relatively small
values of $K_0$~\cite{Ross10}.  With increasing disorder strength the nonmonotonic
dependence on the density is smoothed out and $G(\phi=\phi_0/2)$ becomes equal to $G(\phi = 0)$ and the half-flux quantum period starts to
dominate over the one-flux quantum period. Away from these values of the flux, time-reversal symmetry is broken and the weak antilocalization correction to the conductance gives a $\phi_0/2$ period of the magnetoconductance.

We can estimate the disorder strength at which the crossover between the $\phi_0$ and $\phi_0/2$ period happens with the following argument. The elastic linewidth broadening is given
by~\cite{Ada09}
 \begin{align}
   \frac{\hbar}{\tau} &= \int \frac{d\mathbf{q}d\mathbf{r}}{4\pi} (1-\cos^2\theta_{\mathbf{q}}) \delta(k_F - q)\langle V(0)V(\mathbf{r})\rangle 
   e^{i\mathbf{q}\cdot\mathbf{r}} \\ \notag
   &= \frac{\hbar v}{\xi} \frac{2K_0 I_1[(k_F\xi)^2]}{k_F\xi\exp[(k_F\xi)^2]}
 \label{eq:tau}
\end{align}
with $k_F = \varepsilon/(\hbar v)$. When this broadening is of the order of the mean level spacing $2\hbar v\pi/W$, the oscillations in the density of states that cause the one-flux quantum
period are washed out. For the parameters in Fig.~\ref{fig:K0dep} this gives a crossover value $K_c \approx 0.2$ in reasonable agreement with
the data shown in Fig.~\ref{fig:K0dep}.

In the limit of short and wide wire
the conductance is independent of the flux, and $\delta G_{1/2}$ goes to zero. Intuitively this is
because the particles leave the wire before they have time to go around the circumference of the wire. 
Adapting the calculation of the weak antilocalization correction~\cite{Lee85} to the boundary conditions appropriate to the cylinder
geometry one finds
\begin{equation}
  \delta G = \frac{W}{L} \frac{e^2}{h} \delta \sigma,
  \label{eq:walG}
\end{equation}
with
\begin{equation}
  \delta \sigma = \frac{1}{\pi}\log\frac{L}{\xi}+ 
   \frac{1}{\pi}\sum_{n=1}^{\infty}\cos\frac{4\pi n\phi}{\phi_0}\log(1-e^{-\pi n W/L}),
\end{equation}
up to an $L$- and $\phi$-independent constant. In Fig.~\ref{fig:K0dep} we compare the analytical expression~\eqref{eq:walG} to the
numerical calculation with good agreement.

With a good understanding of the different transport regimes we are now in a position to discuss the findings of the experiment of Peng {\it et~al.}~\cite{Pen10} in the context of our results. A magnetoconductance with a period of $\phi_0$ and a maximum at zero flux is realized in our model at doping large enough that the conductance is larger than $e^2/h$ and in the presence of weak disorder such that Fabry-Perot resonances are washed out but the dynamics are not yet
fully diffusive. It is reasonable to expect the samples to be doped as there is no reason for the Fermi energy to coincide with the Dirac
point. In fact, special measures are required to get them to coincide~\cite{Hsi09}.

Whether the condition of weak disorder is realized is harder to judge,
as very few transport experiments are available. It is, however, clear how one would go about checking experimentally whether this scenario is actually realized. By varying the chemical potential of the surface states the maximum of conductance should shift back and forth between
$\phi = 0$ and $\phi = \pi$ with an amplitude that, depending on parameters, is likely to be largest (close to $e^2/h$) at the Dirac point (cf.~Fig.~\ref{fig:EnergyDependence}). Alternatively, one can change the period of the conductance oscillations from $\phi_0$ to $\phi_0/2$ by increasing the surface disorder (cf.~Fig.~\ref{fig:K0dep}). We expect both methods to be practicable with current experimental techniques.

In summary, we have calculated the flux dependence of the conductance of disordered topological quantum wires in the presence of quantum flux parallel to the
wire, modeling the surface states with a continuous Dirac Hamiltonian. The conductance is found to have the expected $\phi_0/2$ period due to
weak antilocalization away from the Dirac point and with strong enough disorder such that the electron motion is diffusive. A period of
$\phi_0$ is obtained either at the Dirac point at any disorder strength, or away from the Dirac point with weak disorder. While in the former case the conductance always has a maximum at $\phi = \phi_0/2$, in the latter case the maximum can be reached at either $\phi=0$ or $\phi = \phi_0/2$ depending on the doping level.  We hope that these results will aid further transport experiments on the remarkable surface state of topological insulators.

The protection of a mode at $\phi_0/2$ is studied using a fully 3D simulation in a concurrent work~\cite{Zha10}.  We
acknowledge conversations with J.~Cayssol, D.~Carpentier, E.~Orignac, and A.~Vishwanath. This research was supported by the U.S. Department of Energy under Contract No. DE-AC02-05CH11231 (JHB and JEM)
and the Alexander von Humboldt Foundation (PWB).

\end{document}